\documentclass[fleqn,usenatbib]{mnras}

\usepackage{newtxtext,newtxmath}

\usepackage[T1]{fontenc}
\usepackage{ae,aecompl}

\usepackage{graphicx}	
\usepackage{amsmath}	
\usepackage{amssymb}	
\usepackage[caption=false]{subfig}
\usepackage{float}

\title[The FIRST Classifier]{The FIRST Classifier: Compact and Extended Radio Galaxy Classification using Deep Convolutional Neural Networks}

\author[Wathela Alhassan et al.]{
Wathela Alhassan,$^{1,2}$\thanks{E-mail: wathelahamed@gmail.com}
A.R. Taylor,$^{1,3}$
Mattia Vaccari$^{3,4}$
\\
$^{1}$Inter-University Institute for Data Intensive Astronomy, and Department of Astronomy, University of Cape Town, Private Bag X3,\\ Rondebosch 7701, South Africa\\
$^{2}$Institute of Space Research and Aerospace, Khartoum 6096, 111113, Sudan\\
$^{3}$Inter-University Institute for Data Intensive Astronomy, and Department of Physics and Astronomy, University of the Western Cape\\, Private Bag X17, Bellville 7535, South Africa\\
$^{4}$INAF - Istituto di Radioastronomia, via Gobetti 101, 40129 Bologna, Italy}

\date{Accepted XXX. Received YYY; in original form ZZZ}

\pubyear{2018}

\begin{document}
\label{firstpage}
\pagerange{\pageref{firstpage}--\pageref{lastpage}}
\maketitle

\begin{abstract}
Upcoming surveys with new radio observatories such as the Square Kilometer Array will generate a wealth of imaging data containing large numbers of radio galaxies. Different classes of radio galaxies can be used as tracers of the cosmic environment, including the dark matter density field, to address key cosmological questions. Classifying these galaxies based on morphology is thus an important step toward achieving the science goals of next generation radio surveys. Radio galaxies have been traditionally been classified as Fanaroff-Riley (FR) I and II, although some exhibit more complex 'bent' morphologies arising from environmental factors or intrinsic properties. In this work we present the FIRST Classifier, an on-line system for automated classification of Compact and Extended radio sources. We developed the FIRST Classifier based on a trained Deep Convolutional Neural Network Model to automate the morphological classification of compact and extended radio sources observed in the FIRST radio survey. Our model achieved an overall accuracy of $97\%$ and a recall of $98\%$, $100\%$, $98\%$ and $93\%$ for Compact, BENT, FRI and FRII galaxies respectively. The current version of the FIRST classifier is able to predict the morphological class for a single source or for a list of sources as Compact or Extended (FRI, FRII and BENT). 
\end{abstract}




\section{Introduction}

The morphological classification of galaxies is an approach based on grouping them by their visual appearance. Studying radio galaxies on the basis of their morphology allows us to understand the formation and evolution of galaxies and their sub-components as a function of e.g. luminosity, environment, stellar mass and star formation rate over cosmic time \citep{Helfand2015}. Different classes of radio galaxies can be used as tracers of the cosmic environment, including the dark matter density field and galaxy clusters, to address key cosmological questions \citep{Makhathini2015}.

Over the next few years, deep wide-area surveys with new radio observatories such as the Karl G. Jansky Very Large Array (VLA), the Australian Square-Kilometre-Array Pathfinder (ASKAP, \cite{Johnston2007, Johnston2008}), MeerKAT \citep{Jonas2016} and eventually the Square Kilometre Array (SKA, \citep{Brown2015}) will be carried out, and a vast amount of radio images will  become available. Manual inspection of these images will be impractical \citep{Hocking2015}, which motivates developing tools that can automatically analyse them, including automated morphological classification techniques for radio sources.

The radio sky is populated by a variety of compact and extended sources. Compact sources are unresolved sources which have a single non-diffuse component. The overwhelming majority of radio sources at 1.4 GHz fluxes of 1 mJy or so are compact \citep{Banfield2015,Lukic2018}. Extended radio galaxies have been traditionally classified using the Fanaroff-Riley (FR) scheme \citep{1974MNRAS.167P..31F} as FRI and FRII sources. FRIs and FRIIs are distinguished based on the position of low- and high-surface brightness regions in the extended components of the source. FRI sources have smaller separation between the points of peak intensity in the two lobes, namely smaller than half the total extent of the source, and have the highest surface brightness along the jets and core (the edge-darkened FRIs). Conversely, FRII sources have a separation between the two points of peak intensity that is larger than half the total extent of the source and have the highest surface brightness at the edges (edge-brightened FRIIs). Some extended sources also exhibit more complex bent morphologies arising from environmental factors or intrinsic properties and are thus often referred to as bent-tailed sources or simply bent sources. Bent sources can be used to trace clusters at higher redshifts, especially when information from other wavelengths (e.g. optical or X-ray) is not available \citep{Blanton2000,Blanton2001,Blanton2003}. Bent sources can further be classified into two main classes based on how their jets appear. Wide-Angle Tail (WATs) radio galaxies are sources where radio-emitting jets follow a wide C shape due to the dynamic pressure resulting from the host galaxy's rapid motion through the surrounding intracluster medium (ICM) \citep{Sakelliou1999}, located at or close to host cluster's center with higher peculiar velocities \citep{Douglass2007,Douglass2011}. Narrow-Angled Tail (NATs) radio galaxies are sources where the source resides in the cluster's outer regions with larger peculiar velocities and distinguished by its diffuse tail that follows a narrow C shape due to the host galaxy’s rapid motion through the ICM, at higher resolution its tail can often be seen splitting up into two tails \citep{1989MNRAS.238..357O}. NATs and WATs are also called - due to their bright head - Head Tail (HT) sources in the literature \citep{Proctor2011}.


The application of artificial neural networks to the problem of optical galaxy morphology classification has been the subject of active work since the early nineties
\citep{StorrieLombardi1992,Lahav1995,deLaCalleja2004}. 
The application of Convolutional Neural Networks (CNNs) to computer vision goes back to 1998, achieving good results for handwritten digit classification \citep{LeCun1998}. With the development of computing technology, CNNs have recently shown state-of-the-art performance on image classification \citep{Krizhevsky2012}. Important improvements have been achieved in visual recognition of many categories \citep{Jiang2016}. In astronomy, projects such as Galaxy Zoo \citep{Lintott2008} have thus generated strong interest in applying convolutional neural networks to visually classified galaxy samples \citep{Dieleman2015}.



Whilst most of machine learning exploitation in astronomy has been done on optical data, little work has been done on the morphological classification of radio galaxies. Unsupervised radio source classification has been performed using the Self Organizing Kohonen Map dimensionality reduction technique \citep{Polsterer2011,Polsterer2015}, which combines and sorts similar sources into classes and produces a single template representation of every class.

An application of Convolutional Neural Networks to extended radio galaxy morphology was presented by \cite{Aniyan2017}, where they classified extended radio sources into three types, FRI, FRII and Bent, with an average precision of 91\%, 75\% and 95\% respectively. In this work we have extended the problem to include compact sources using a similar dataset with different model structure and data augmentation techniques.

Following the developments in optical galaxy morphology classification, the Radio Galaxy Zoo project \citep{Banfield2015} has recently engaged many citizen scientists to identify the morphological type of radio sources and determine their host galaxy by combining infrared and radio observations. However, their classification scheme is based on the number of components of extended radio sources and does not lend itself to be interpreted in terms of FRIs, FRIIs and bent sources. However, \cite{Lukic2018} have applied convolutional neural networks to the classification of sources according to this scheme and achieved a final test accuracy of 94.8 per cent on Radio Galaxy Zoo Data Release 1.

The main purpose of this work is to automate the morphological classification of compact and extended radio sources (three classes FRI, FRII and BENT (WAT and NAT)) by developing a classifier that uses a trained deep Convolutional Neural Network (CNN) model to generate accurate and robust predictions. We developed our CNN model based on the \textbf{Keras} Deep Learning framework \citep{chollet2015keras} using the \textbf{TensorFlow} \citep{tensorflow2015-whitepaper} backend. TensorFlow is an open source software library for numerical computation developed by the Google Brain Team within Google’s Machine Intelligence research organization. 

This paper is organized as follows. In Section 2 we present our Dataset Characteristics. Preprocessing and Data Augmentation are presented in Section 3. Our CNN and our Network Architecture are described in Section 4 and Section 5 respectively, while Section 6 details the model performance evaluation. Sections 7 and 8 are devoted to the online FIRST Classifier and to our conclusions respectively.

\section{Radio Galaxy Catalogue}

We constructed our sample of extended radio sources from three catalogues, each of which contains the source coordinates and their classification label.

For FRIs we used the FRICAT catalogue by \cite{Capetti2016}, which merges data from NRAO VLA Sky Survey (NVSS) \citep{Condon1998}, Faint Images of the Radio Sky at Twenty Centimeters (FIRST) \citep{Becker1995} and the Sloan Digital Sky Survey (SDSS, \cite{York2000}). FRICAT consists of 219 FRI galaxies with redshifts $\leq$ 0.15. All the sources included here have an edge-darkened radio morphology with radius extending larger than 30 kpc from the host.

For FRIIs we used the FRIICAT catalogue \citep{Capetti2017}, which like the previous catalog contains samples that were obtained by merging observations from NVSS, FIRST and NVSS. FRIICAT consists of 122 FRII galaxies and contains sources that have an edge-brightened radio morphology with redshifts $\leq$ 0.15.

FRICAT and FRIICAT are essentially a subset of the catalogue of 18,286 radio sources built by \cite{BestHeckman2012} as a response to the shortage of FRI sources in the literature and to study the main properties of FRI and FRII galaxies based on their spectroscopic classification. They used FIRST images for their morphological classification at FIRST angular resolution ($5"$), and all sources were chosen to have a redshift $z \leq 0.15$ to make sure they are well resolved. For a source to be classified as an FRII, it must have emission peaks at least 30 kpc from the optical host center. Each of the three authors of FRIICAT performed this inspection for each source independently and they only included sources where at least two of them agreed that it was a FRII.

In order to create a reliable catalogue of bent sources, objects in the FIRST catalogue were examined and separated, using different pattern recognition techniques and visual inspection, into lower-count membership (singles, doubles, triples) groups  and higher-count membership (more than three members) groups by \cite{Proctor2011}. They classified the groups with four and more members (7016 groups) into different bent types including WAT, NAT and Compact. To define our bent-tailed (hereafter BENT) sample we only used the confirmed WATs and NATs, which amount to $192$.

For compact (hereafter COMP) sources, we made use of the Combined NVSS-FIRST Galaxy catalogue (CoNFIG, \cite{Gendre2009,Gendre2010}), which include new VLA observations, optical identifications and redshift estimates of Compact, FRI and FRII sources. The catalogue consists of 859 sources over 4 samples (CoNFIG-1, 2, 3 and 4 with flux density limits of $S_{1.4GHz}$ = 1.3, 0.8, 0.2 and 0.05 Jy respectively). It is 95.7{\%} complete in radio morphology classification and 74.3{\%} of the sources have redshift data. All the sources smaller than 3~arcsec were classified as compact sources. We also made use of FRIIs sources in this catalogue and added them to the FRIICAT sources.

The number of sources in our sample is summarised in Table ~\ref{tab:no_sample} 

\section{Image Pre-Processing and Data Augmentation}

The FIRST survey \citep{Becker1995} mapped approximately one quarter of the sky at 1.4 GHz with a 5" angular resolution and a 1 mJy/beam sensitivity. The survey covers 10,575 square degrees of sky with 8,444 square degrees in the north and 2,131 square degrees in the south. The survey produced a 21 cm source catalogue with flux densities, subarcsecond positions, and morphological information for about a million sources. Both the northern and southern areas were chosen to coincide approximately with the area covered by the Sloan Digital Sky Survey (SDSS, \cite{York2000}).

We retrieved the FIRST images for the catalogue described in the previous section via the online FIRST image archive as FITS files. Images were first cleaned, rescaled and cropped to reduce the dimensionality of the input and then saved as PNG files. It was useful to crop the images because the object of interest is in the middle of the image with a large amount of sky background, and to reduce the number of features to be extracted. We then rescaled the images to speed up training, with little to no effect on predictive performance. Images were cropped from $300 \times 300$ pixels to $150 \times 150$. For a small number of images, where an extended source was either extremely large or not perfectly at the center of the image, the cropping operation removed part of the sources. We thus removed 84 sources from our dataset.

For the cleaning process we used the same method adopted by \cite{Aniyan2017}, where all the pixels values below $3\sigma$ were removed (set to zero) in order to subtract the background noise. Due to the small number of labelled images, artificial images were created by flips and rotations, to generate sufficient data for training our model. Every labelled image was rotated by a random angle, then flipped along the x-axis to produce the artificial images. Flips and rotations do not increase the topological information contained within the data, but obviously alter the orientation of the object.

\begin{figure}
\centering
\includegraphics[scale=0.55]{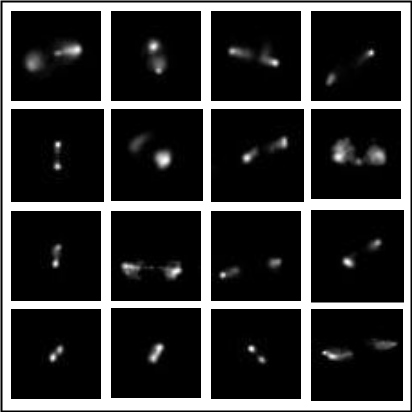}
\caption{FRII galaxies sample images. }
\label{fig:fr2}
\end{figure}

\begin{figure}
\centering
\includegraphics[scale=0.6]{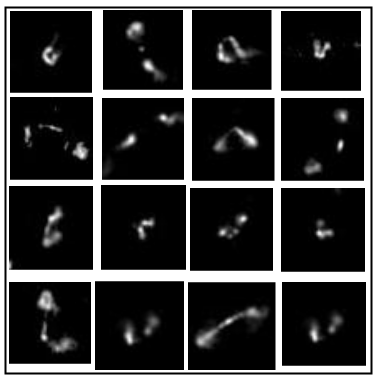}
\caption{BENT galaxies sample images (WATs and NATs).}
\label{fig:bent}
\end{figure}

\begin{figure}
\centering
\includegraphics[scale=0.55]{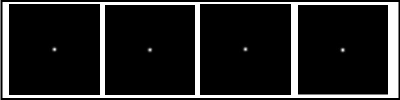}
\caption{Compact galaxies sample images.}
\label{fig:comp}
\end{figure}

These operations have been carried out on every image from our original dataset (837 images). Figure \ref{tab:no_sample} details the number of our original images, Flipped/Rotated version of them and the proportions of training, testing and validation of our final dataset. Sample images of FRI, FRII, BENT and COMP sources are shown in Figure \ref{fig:fr1}, \ref{fig:fr2}, \ref{fig:bent} and \ref{fig:comp} respectively, after being pre-processed.
The augmented dataset, along with the corresponding class labels, were then used to train our CNN model for the purpose of Compact and Extended radio galaxies morphology classification.

\section{Convolutional Neural Networks (CNNs)}

The Convolutional Neural Networks (ConvNet; \cite{Fukushima1980}) is a type of feed-forward neural network model -- meaning the output from one layer is used as input to the next layer -- for deep learning. CNNs consists of multilayer structure, namely convolutional layers (which act as a filter for the input images to extract features), followed by an activation non-linearity function such as $tanh$, $sigmoid$ or $ReLU$, then pooling layers, which are vector to scalar transformations that operate on local regions of an image to generate a representative value of the pixels in that region. Average pooling computes the average of pixels in a region, while max pooling uses the value of the pixel with the highest intensity in the region, optionally followed by fully connected layers \citep{Agarap2017}.

In general the architecture of CNNs is designed to take advantage of the 2D structure of an input image (or other 2D input such as a speech signal) \citep{Goodfellow-et-al-2016}. Convolutional layers are essentially made up of neurons that receive inputs. Each neuron is connected locally to its inputs from the previous layer. The inputs are each assigned a random weight, and a dot product is performed. The scalar output is then passed through a non-linear activation function.

The input to the convolutional layer is an image of size $m \times m \times c $ where $m$ is the width and height of the image and $c$ is number of bands (e.g. a gray-scale image has 1 band and an RGB image has 3 bands). The convolutional layer has filters of size $n \times n \times g$ (where $ n < m$ and $g \leq c$) and the initial values of filters (weight matrices) are user-defined. Each weight element is convolved with the image to produce k channels called feature maps of size $m \times n+1$, each feature map will be sub-sampled with a pooling layer (either max, min or average) over contiguous regions (usually $2 \times 2$ for small images or $3 \times 3$ for larger one) \citep{Krizhevsky2012}. The output k-th feature map $Z$ of a single neuron is a result of a non-linear transformation; and can be mathematically represented as:

\newpage
\begin{figure}
\centering
\includegraphics[scale=0.55]{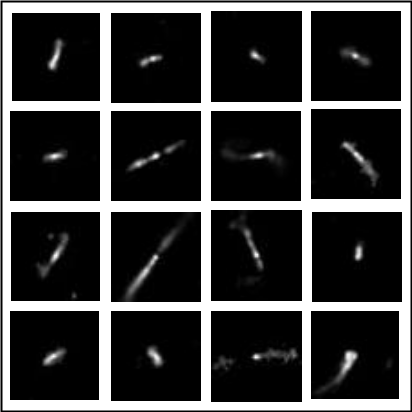}
\caption{FRI galaxies sample images }
\label{fig:fr1}
\end{figure}
\begin{figure}
\centering
\includegraphics[scale=0.5]{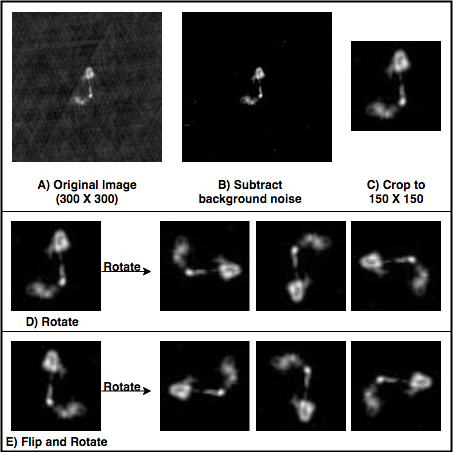}
\caption{Pre-processing steps of the images}
\label{fig:notable}
\end{figure}
\begin{table}
\centering
\begin{tabular}{cccccc}
\hline
  Type & Original Sample & F/R& Train & Test& Val\\
  \hline
  COMP & 121 & 5,445 & 3,445 & 1000 & 1000 \\ 
  FRI & 201 & 5,445 & 3,445 & 1000 & 1000 \\
  FRII& 338 & 5,545 & 3,445 & 1000 & 1000 \\
  BENT & 177 & 5,445 & 3,445 & 1000 & 1000 \\
  \hline
  Total & 837 & 21,780&13,780& 4000 & 4000 \\
  \hline
\end{tabular}
\caption{Number of: original sample images, flipped/rotated version images (F/R), training, testing and validation images.}
\label{tab:no_sample}
\end{table}

\clearpage
%

\begin{equation}
Z^{k} \mbox{=}  f( w^{k} * x) = f(\sum_{i^{'}=1}^{n} \sum\limits_{ j^{'}=1}^{n}  w_{i^{'},j^{'}}^{k} x_{i+i^{'},j+j^{'}}+ b^{k})
\end{equation}

where $x$ is the activations of the input neurons connected to the neurons ($i,j$) in the following convolutional layer, $w$ is an $n \times n$ weight matrix of the convolutional filter, b is the bias, $*$ is the convolution operator and $f()$ is a non-linear activation function, usually ReLU, Tanh, sigmoid or softmax.
The $ReLU$ activation function \citep{Hahnioser2000} sets negative values to $0$ and returns only the positive values, frequently used for hidden layers; mathematically this can be represented as:

\begin{equation}
 y_i =  \left\{ \begin{array}{rcl}
 x_i & \mbox{if} & x_i\geq 0 \\
 0  & \mbox{if} & x_i < 0
\end{array}\right.
\end{equation}

The $TanH$  activation function, squashes a real number $\in R$ to the range $\in [-1,1])$. The $softmax$ \citep{10.1007/978-3-642-76153-9_28},  converts the score of each class to probabilities (Convert scores $\in R$ to probabilities $\in [0,1])$ where the final output prediction is the highest probability class, $softmax$ mathematically can be represented as:

 \begin{equation}
P(class) =   \frac{exp^{Z_{i}}}{\sum\limits_{i} exp^{Z_i}}
 \end{equation}
where Z is the the output (score) from the previous layer. Similarly, $sigmoid$ activation function takes a vector of weight and produces scalar output in the range between $0$ and $1$. The whole network describes the non-linear mapping between raw image pixels and their class scores.
\section{Model's Network Architecture}

\begin{figure}
\centering
\includegraphics[scale=0.70]{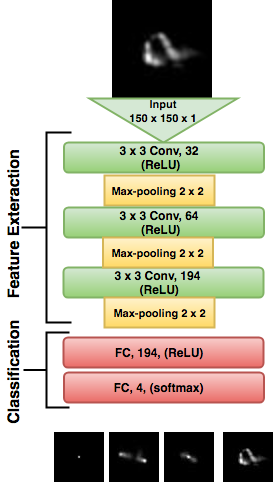}
\caption{ Deep CNN model architecture. In this Figure each convolutional layer (Conv) is followed by ReLU activation function and Max-pooling function. The number in each box represents the number of channels in the corresponding feature map.}
\label{fig:model}
\end{figure}

The radio images at our disposal can be described by a small number of features because the number of useful pixels after the cleaning process is rather limited. Based on that, we assumed a simple network architecture (i.e. a small number of convolutional layers) will solve our problem. We constructed different models with different numbers of convolutional layers(2 to 10) and different types of activation functions, we only described the structure of the best performance network for the classification of the radio galaxies images here, which illustrated in Figure \ref{fig:model}. The network consists of five trainable layers. The first convolutional layer filters the $150 \times 150 \times 1$ input image (gray image) with $32$ square filters of size $3\times3$. The second Convolutional layer filters the output of the first one with $64$ filters of size $3\times3$ and the last convolutional layer filters the output of the second layer with $194$ filters of size $3\times3$.  A $ReLU$ activation function \citep{Hahnioser2000} was applied to the output of the all three convolutional layers.
 
 The output of all the convolutional layers were sub-sampled with Max-pooling with of size $2\times2$, the last Max-pooling was followed by $2$ fully-connected (FC) layers. The first one had 194 channels with $ReLU$ activation function and a $Dropout$ function \citep{Hinton2012}, Drop out is a regularisation technique which is used to avoid over-fitting by setting randomly the output of the previous layers neurons to $0$, to force the model to learn robust features rather than relying on the presence of same features each time. The second FC layer performs the classification, where its output was fed to the $softmax$ activation function.

Compared to other activation functions, $ReLU$ has been proven to accelerate the convergence of stochastic gradient descent by a factor of $6$ because of its linear non-saturating form \citep{Krizhevsky2012, Nair:2010:RLU:3104322.3104425}, $ReLU$ also has a simple (cheap) operation  compared to the exponential in $Sigmoid$ case and gives better performance than $TanH$ \citep{Glorot2011}. In the output layer, using $softmax$ and $sigmoid$ as activation function is a common practice, unlike $softmax$, in the case of $sigmoid$ when the neuron's activation saturates at either tail of $0$ or $1$, the backpropagation algorithm fail at modifying its parameters because of the gradient value at these regions is almost zero \citep{Goodfellow-et-al-2016}, based on that we made use of the $softmax$ in the output layer.

The model was trained on radio galaxies images comprising of 4 classes  for $400$ epochs. Weights and biases were initialised using Xavier initialisation. The model has $12,325,792$ learnable parameters, Table \ref{tab:hpara} show the hyper-parameters used to train our model.

\begin{table}
\centering
\begin{tabular}{cc}
\hline
   Hyper-parameters & Values\\
  \hline
  Batch Size&128 \\ 
  Dropout Rate & 0.5\\
  Learning Rate &  0.0001  \\
  Epochs&   400  \\
  \hline
\end{tabular}
\caption{Hyper-parameters used to train the CNN model}
\label{tab:hpara}
\end{table}

\section{Model evaluations}
\subsection{Metrics and Quantitative Experimental Results}
\begin{figure}
\centering
\subfloat{\includegraphics[scale=0.40]{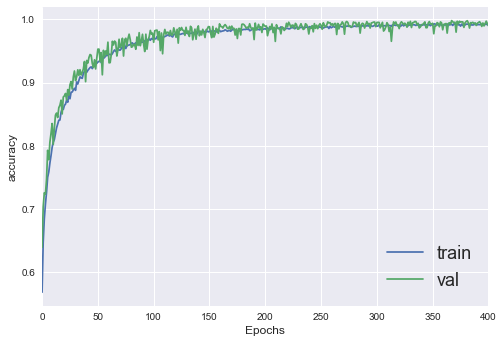}}\\
\subfloat{\includegraphics[scale=0.40]{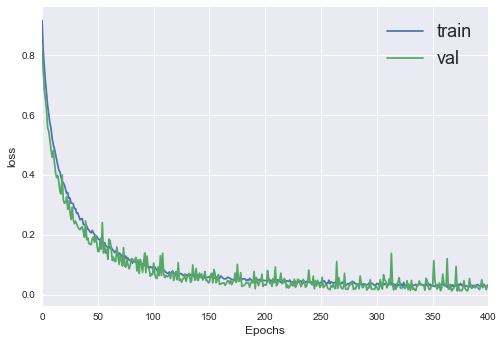}}
\caption{Training and testing  learning curves of our CNN model with respect to the number of epochs in the training phase. Top: Training and testing accuracy curves, Bottom: training and testing loss curves .}
\label{fig:curves}
\end{figure}

For measuring the performance of the deep learning model, the learning curve is generally used, which provides a quantitative measure of the performance of the model on the training and validating dataset, in terms of accuracy and loss, with respect the to the number of epoch \citep{Urry2012}. Losses and accuracy curves for the training and validation datasets were plotted with respect to the number of epochs(iterations). As shown in Figure \ref{fig:curves} the accuracy increases while the loss decreases through the training process until convergence, The Classification  achieved an overall accuracy of $0.97$ and a loss of $0.09$ for Training and Validation. Relatively simple CNN architecture (only three convolutional layers) performs better in this case.

In order to assess how accurately our model is able to predict the different morphology classes, the precision (P), recall (R) and F1 scores \citep{Fawcett2006} were calculated using our test dataset, based on the number of true positive (TP), false positive (FP) and false negative (FN) classifications as given below:
\begin{equation}
recall = \frac{TP}{TP +FN}
\end{equation}

\begin{equation}
Precision = \frac{TP}{TP+FP}
\end{equation}

\begin{equation}
F1 score = 2 \times \frac{Precision \times Recall}{Precision + Recall}
\end{equation}

where:
\begin{itemize}
\item[$\bullet$] true positive (TP) is when the source for instance is predicted as FRI and it is actually FRI.

\item[$\bullet$] false positive (FP) is when the source is predicted for instance as FRI and it is actually not FRI.

\item[$\bullet$] false negative (FN) is when the source is predicted for instance as not FRI and it is actually FRI.
\end{itemize}

Generally, the recall is the best measure to check for over-fitting,  while the precision provides a quantitative measure of the correctly classified sources. Recall and precision are also called reliability (or sensitivity) and completeness respectively \citep{Hopkins2015}. The higher value of recall and precision the better performance of the model \citep{Ivezic:1662945}. F1 score is a weighted average of the recall and the precision, which is a good measure for the classification, higher value of F1 score means better performance of the classification. Table ~\ref{tab:report} shows the classification report where the precision, recall and F1 score were calculated for each of our 4 classes. Excellent results were achieved for precision, recall and F1 score  for all source classes. In more detail, the best performance was achieved for FRI sources with a recall of $100\%$, Followed by Compact and Bent sources with $98\%$ and FRII sources with $93\%$, In terms of precision, Compact and FRI achieved $98\%$ while Bent and FRII got $96\%$. 

\begin{figure}
\centering
\includegraphics[scale=0.60]{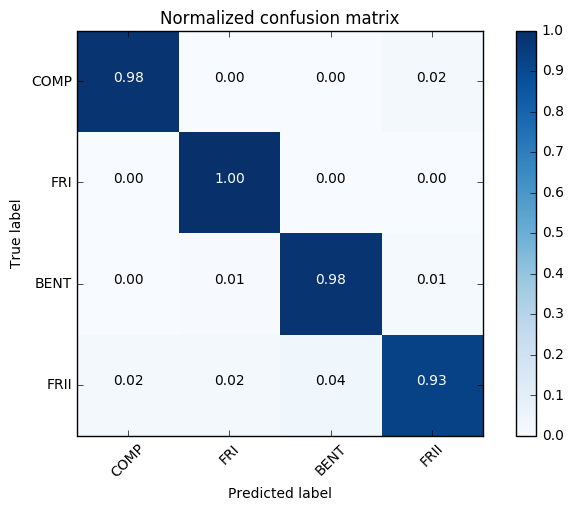}
\caption{Normalized Confusion Matrix on our testing dataset, where the true labels at $Y$ axis and Predicted labels at the $X$ axis, the blue boxes at the diagonal represents the recall values .}
\label{fig:cm}
\end{figure}

The Confusion Matrix (CM) is used to visualize the performance of Machine Learning algorithms \citep{Stehman1997}, its stability was proved generally in the case of multi-class classification tasks \citep{Machart2012}, CM is essentially a matrix where each row represents the instances in a predicted class ($X axis$) while each column represents the instances in an actual class ($Y axis$), the classes ("actual" and "predicted") on both axis are identical which enables comparison with the ground truth, and makes it easier to interpret and see if the system confuse the classes. Figure \ref{fig:cm} shows the normalized confusion matrix of our testing dataset, where the recall, TP, FP and FN values were normalized to 1 and plotted for compact, FRI, FRII and BENT sources. 

\subsection{ Qualitative experimental results}
\begin{figure}
\centering
\includegraphics[scale=0.6]{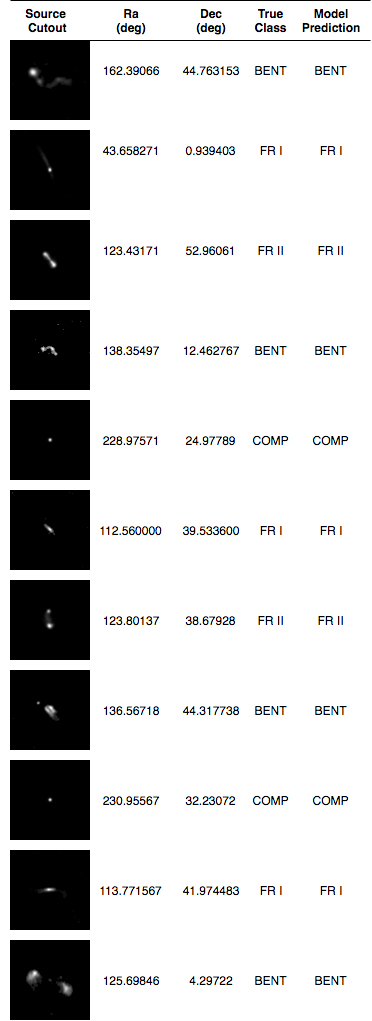}
\caption{Comparison between our model predictions and ground truth of random samples from FIRST Images}
\label{fig:comparison1}
\end{figure}

To qualitatively evaluate the effectiveness of our CNN model in comparison with the ground truth (True labels), image samples classified by our CNN models were visualized as shown in Figure \ref{fig:comparison1}, along with: predicted class, actual class(ground truth) and the coordinates (Right Ascension and Declination). All the samples for the classification were $100\%$ accurate and no misclassification were spotted, which confirms the quantitative results that our model has achieved.

\begin{table}
\centering
\begin{tabular}{ccccc}
\hline
  Type & precision & recall & f1-score & support\\
  \hline
  COMP & 0.98 & 0.98 & 0.98 & 1000\\ 
  FRI  & 0.98 & 1.00 & 0.99 & 1000\\
  BENT & 0.96 & 0.98 & 0.97 & 1000\\
  FRII & 0.96 & 0.93 & 0.95 & 1000\\
  \hline
  avg/ total & 0.97 & 0.97 & 0.97 & 4,000\\
  \hline
\end{tabular}
\caption{The table shows the classification report of the testing dataset, where the precision, recall and F1 score were calculated for Compact(COMP), FRI, FRII and BENT sources, the number of sources for each class is shown in the support column.}
\label{tab:report}
\end{table}

\section{FIRST Classifier}

We developed the FIRST classifier based on our CNN model. The current version of the classifier has access to all the sources available at the final release of the FIRST survey (946,432 sources) \citep{Helfand2015}, which retrieves a postage-stamps from the FIRST archival data using the given coordinate of the source. Each postage-stamp has a well resolved single source which is the input to the model. The classifier consists of two main parts: the pipeline part which features an automatic search for radio sources from FIRST survey archived images \citep{Becker1995} (which the name FIRST comes) remotely using Virtual Observatory tools (PyVO) \citep{Graham2014}. PyVO provides access to Virtual Observatory services and archived data from different surveys remotely using python. 

\clearpage


\clearpage
%


\begin{figure}
\centering
\includegraphics[scale=0.5]{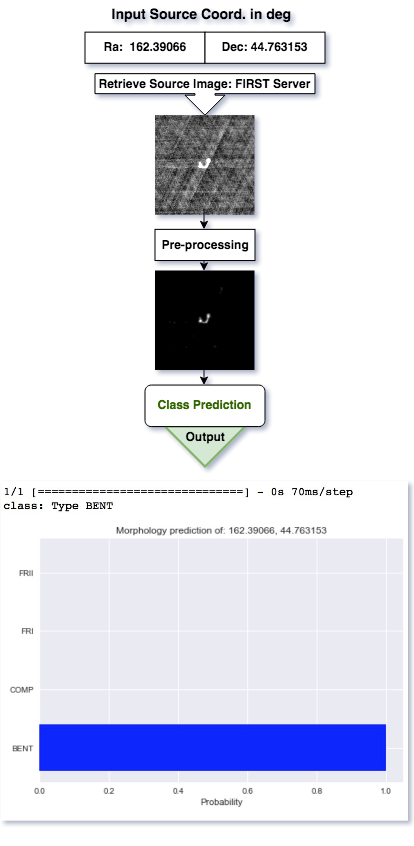}
\caption{Schematic overview of FIRST Classifier.}
\label{fig:clas}
\end{figure}

As shown in Figure \ref{fig:clas} the classifier performs the classification through three steps: (i) Input: FIRST classifier has an input:  coordinates of single radio source or a list of coordinates of multiple radio sources (csv file).   Then retrieves cut-out images of the input coordinates from the FIRST archived images, (ii) Clean the retrieved image by subtracting the background noise and cropping the inner part of the image to a size of  $150 \times 150$ , (iii) Feed the source's image to the model to predict its morphology type. This model as mentioned previously is trained to recognize Compact and Extended radio sources.  For Extended sources it classifies into three morphology types: FRI, FRII and BENT, (iv) Output: if the input to the classifier was a coordinate of a single source, output will contain the predicted morphology type(corresponding to the highest probability), probabilities plot of the classification and a direct link to download the FITS file cut out of the target. If the input was a list of sources's coordinates, output will be a csv file containing 4 columns: Coordinates (RA and DEC), Predicted class, Highest probability, Link to download the cut-out FITS file. If the target source was not found a "data not available error" will be raised, since the image retrieving process is on-line, an error of "Time out please re-run again" will appear in case of weak Internet connection.
\section{Conclusions} 
Upcoming surveys with radio observatories like the VLA, ASKAP, MeerKAT and eventually the SKA will generate very large images containing vast numbers of radio galaxies. Different classes of radio galaxies can be used as tracers of the cosmic environment, including dark matter density field, to address key cosmological questions. Manual inspection of these images will be impractical, which motivates developing tools that can automatically analyse them, this includes developing an automatic morphological classification of radio sources.

A CNN model with only three convolutional layers, with batch size of $128$, $400$ Epochs and Learning Rate of $0.0001$ was presented for classifying Compact and Extended radio galaxies observed as part of the  the FIRST survey. Our model achieved excellent results with a recall, precession and F1-Score of $97\%$. Based on this model an automatic classifier for radio sources imaged by the FIRST radio survey was developed and presented. The FIRST Classifier is an on-line system for automated classification of FIRST sources, which works well for Compact and Extended Radio Galaxies available on the FIRST image archive, in very rare cases, the double sources in FIRST images might be extended over larger size of the cut-out (> $150 \times 150$), which implies that, after performing the preprocessing, some of the source might be lost which will results in inaccurate classification. This issue will be solved in the newer coming version of the classifier by training the model on larger size of cut-outs (eg. $200 \times 200$), development will also focus on improving the current version of the classifier to be able to handle large images each containing multiple sources.

The FIRST Classifier is publicly available at \url{https://github.com/wathela/FIRST-CLASSIFIER}, we have tested it for correctness and robustness, and only basic computer skills are required to use the classifier as a command line utility. Researchers in radio astronomy are encouraged to make use of the classifier and provide us with their feedback about its performance. 

Future work will focus on improving the classifier to handle data from MeerKAT, VLASS \citep{Myers2018}, ASKAP and finally SKA. For this task, a cut-out generator will be developed, and a manual morphological classification of samples from VLASS, EMU \citep{Hopkins2015}, MIGHTEE \citep{Jarvis2017} and MeerKLASS \citep{Santos2017} images will be performed to support the training dataset.  

\section*{Acknowledgements}

The National Radio Astronomy Observatory is a facility of the National Science Foundation operated under cooperative agreement by Associated Universities, Inc.


\bibliographystyle{mnras}
\bibliography{first_classifier} 





\bsp	
\label{lastpage}
\end{document}